\documentclass{aa}
\usepackage{psfig}
\usepackage{natbib}

\begin{document}
\title{A nanoflare heating model for the quiet solar corona} 
\author{Urmila Mitra-Kraev \and Arnold O.~Benz}
\mail{urmila@astro.phys.ethz.ch\\benz@astro.phys.ethz.ch}
\institute{Institute of Astronomy, ETH-Zentrum, CH-8092 Z\"urich, Switzerland}
\date{Received 29 January 2001 / Accepted 9 April 2001}
\titlerunning{Heating the quiet solar corona}
\authorrunning{U.~Mitra-Kraev \& A.~O.~Benz}
\abstract{
The energy input into the lower solar corona by flare evaporation events has 
been modeled according to the available observations for quiet regions. The 
question is addressed whether such heating events can provide the observed 
average level of the coronal emission measure and thus of the observed flux 
of extreme ultraviolet (EUV) and X-ray emission without contradicting the 
observed average power spectrum of the emission measure, the typical emission 
measure variations observed for individual pixels and the observed flare 
energy distribution. As the assumed flare height influences the derived flare 
energy, the mathematical foundations of nanoflare distributions and their 
conversion to different height assumptions are studied first. This also 
allows a comparison with various published energy distributions differing in 
height assumptions and to relate the observations to the input parameters of 
the heating model. An analytic evaluation of the power spectrum yields the 
relationship between the average time profile of nanoflares (or microflares), 
assumed to be self-similar in energy, and the power spectrum. We find that 
the power spectrum is very sensitive to the chosen time profile of the 
flares. Models are found by numerical simulation that fit all available 
observations. They are not unique but severely constrained. We concentrate on 
a model with a flare height proportional to the square root of the flare 
area. The existence of a fitting model demonstrates that nanoflare heating of 
the corona is a viable and attractive mechanism.

}
\maketitle

\section{Introduction} 

Many mechanisms have been proposed in the past for heating the solar corona
\citep*[e.g.~reviews in][]{urp91}.  
Most popular at present are models based on either the release of magnetic
energy or the dissipation of waves. Both scenarios were proposed soon
after the discovery of the high coronal temperature by Edl\'en and Grotrian in
the late 1930s. As the magnetic energy dominates the coronal energy density,
the release of free magnetic energy, possibly present in the corona in the
form of electric currents, is a very suggestive idea. However, it was soon
realized that the electric resistivity caused by collisions between electrons
and ions is extremely low in the thin corona, making this mechanism
insufficient. Therefore, several wave heating mechanisms have been
proposed. Most recent theories involve Alfv\'en waves which can explain both
the preferred heating of regions with high magnetic fields and solar wind
acceleration \citep[e.g.~][]{martu97}. 

\citet{gold64} seems to have been the first to suggest that the building up of
free magnetic energy in the corona must be dissipated by flares when the
resistivity becomes finite by some instability. The idea was further developed
by \citet{levine74}, \citet{parker83}, \citet{heypri84} and
others. \citet{cargill94} pointed out that impulsively 
heated loops would cool by conduction and radiation with observable
results. The general view of these studies was confined to the problem of 
releasing magnetic energy in the corona with the goal to keep it hot.

This is not supported by recent observations. Deep exposures of the soft X-ray
emission in a quiet region by Yohkoh/SXT revealed that the emission measure of
the corona in certain pixels was not constant over the observing time of about
one hour \citep{kbab97}. It suggests that the material content of the
low corona, where most of the soft X-ray emission originates,
varies. More sensitive observations of EUV lines from
the corona show variability of the majority
of pixels (85\%), including some in the faint intra-cell regions
\citep{benkru98,bcm98}. The increases of the emission 
measure are not adiabatic compressions and can only be interpreted as
additions of new material into the corona. The heated material subsequently seems to cool on a
time scale of the order of 15 minutes. Thus the coronal material in the lower
corona appears to be not heated, but rather continuously replaced. We may add
here, however, that a heating process could still be hidden in the
quasi-stationary background of the emission measure.  

Some large events have recently been tested by \citet{bkgb00} for their
relation between increases in emission measure and temperature. The authors
conclude that this relation corresponds more to an impulsive heating of the upper
chromosphere and subsequent expansion rather than a heating at the top of a
coronal loop followed by conductive readjustment. Thus the observed increases
of the coronal emission measure may be interpreted as "chromospheric evaporation"
similar to regular flares in active regions. \citet{benkru99a} and
\citet{kruben00} find very little difference between flares and the observable
properties of the emission measure increases, and thus confirm 
the latter as real nanoflares. In the following, we will use the term
nanoflare, first introduced by \citet{parker83} as a theoretical concept, for any brightening of the quiet corona with energy below approximately 
$10^{26}$\,erg.   

\citet{kruben98} estimated the total energy input by the emission measure
increases to be 16\% of the calculated total radiative loss of the observed
region. This number depends strongly on the sensitivity of the instrument and
some model parameters. In particular, the effective line-of-sight thickness of
the coronal plasma (or height for observations in the center of the disk)
cannot be measured and must be assumed. The distribution of the events in
energy is therefore still controversial. Most observers report a power law
shape, but widely disagree on the exponent, which ranges from $-1.45$
\citep[measuring radiation loss in Fe XII]{bercle99} to $-2.59$
\citep[measuring emission measure increases]{kruben98}. \citet{benkru00}
reported agreement between their EIT based analysis and studies based on TRACE data by \citet{parjup00} and \citet{atnetal00}, if the same
method is used. For a flare model with a height proportional to the square
root of 
the flare area, a simultaneous peak time within 2 minutes over the flare area
and no further flare selection, all three investigations yield a power law
index in the range between $-2.0$ and $-2.4$, the most likely new EIT value being
$-2.3$. Clearly, nanoflare heating of the quiet corona is strongly supported by
the above observations on total energy input and the relatively steep slope of
the energy distribution, which suggests that the smallest flares
contribute most to the heating.

Here we address the question of whether nanoflares can account for all of the
observed properties of the emission measure in the quiet corona, including
{\sl (i)} the general appearance of individual pixels' emission measure time
dependence, including background and nanoflares, {\sl (ii)} the absolute value of
the quasi-steady emission measure (equivalent to the total radiative loss),
and {\sl (iii)} the average Fourier spectrum of the emission measure in time,
reported to be a power law with an exponent of $-1.76$ \citep{benkru98}. 

First, we derive
the conversion of flare frequency distributions in energy for 
different assumptions, in particular the conversion between distributions of flares, where pixels have been grouped into events, and distributions of single pixels, as well as the
conversion between different height models (Sect.~\ref{distribution}).
In Sect.~\ref{simulation} we simulate the emission measure fluctuations in the
quiet corona by 
taking the observed flare frequency distribution in energy, making certain
assumptions on the spatial and temporal shape of the events and 
extrapolating them to the smallest energies needed to explain the total 
radiation loss from the corona. By choosing an adequate time profile for each   
flare, we can simulate the time dependence of the emission measure for an
arbitrary pixel. We also obtain an averaged power spectrum.
In Sect.~\ref{powerspectrum} we calculate the expected value of the power
spectrum analytically and in Sect.~\ref{discussion} the results are
discussed. 
Section \ref{conclusion} summarizes the conditions needed for
the nanoflare model to reproduce the observed features of the variations and 
quasi-steady background of the quiet corona.


\section{Flare Distributions} \label{distribution}  
 
The observations that have been made of the quiet Sun to determine the energy  
input into the corona are based on a grid of many  
pixels in a large field of view. For each pixel and time interval, the 
emission  
measure is estimated \citep[cf.][]{benkru98,parjup00}. Neighboring 
and simultaneously occurring  
events are combined and interpreted as one flare.  
The thermal energy of an emission measure increase can be estimated 
\citep{benkru98}  
\begin{equation} \label{E}  
  E \approx 3 k_B T \sqrt{\Delta{\cal M} A h},  
\end{equation}  
where the emission measure increase $\Delta{\cal M}$, the flare area $A$ and  
the temperature $T$ are observed quantities and $k_B$ is the Boltzmann  
constant. Equation (\ref{E}) assumes that the newly injected material is much 
denser than the  pre-flare density in the loop. As pointed out by
\citet{bkgb00}, this assumption yields an upper limit to the energy. On the other
hand, the flare energy release may also directly heat the coronal material,
whose energy is not included in Eq.~(\ref{E}). In the following, we will use
Eq.~(\ref{E}) as a rough estimate for the thermal energy input by a nanoflare.  
The height (thickness) $h$ of the flare loop lies in the line of  
sight and is therefore not easy to determine. We will use it here as a free 
model parameter that also includes the filling factor. Thus $h$ denotes the 
{\sl effective} height.   
 
All observers agree that the flare frequency distribution function 
depends on the energy as a power law   
\begin{equation} \label{f(E)}  
  f(E) = f_0 E^{-\delta},  
\end{equation}  
where the flare distribution is defined as the number of flare events within a 
given energy interval divided by the total observed area, total observation 
time and energy interval. Similarly, the flare distribution in area was 
reported to be a power law \citep{atnetal00,ali00}. Supported by these observed 
distributions, it is assumed in the following that flares are 
self-similar in thermal energy. Thus on average, a flare is completely 
characterized by its energy. This assumption will later be tested against 
observations.  

Let us therefore write the emission measure increase, the flare area and the
flare height, each averaged for a given thermal energy, in terms of some power
of the energy   
\begin{eqnarray}  
  \Delta{\cal M} &=& c\cdot E^\gamma \label{M(E)} \\  
               A &=& a\cdot E^\alpha \label{A(E)} \\  
               h &=& b\cdot E^\beta  \label{h(E)}.  
\end{eqnarray}  
Where no confusion is possible, we will use the same symbols for averaged 
entities, such as $\Delta{\cal M}$, and measured quantities in individual 
events. Equations (\ref{M(E)})-(\ref{h(E)}) together with Eq.~(\ref{E}) imply
the two conditions  
\begin{eqnarray}  
  \alpha + \beta + \gamma &=& 2  \label{cond1}\\  
  abc (3k_BT)^2           &=& 1. \label{cond2}  
\end{eqnarray}  

It is seen immediately from Eq.~(\ref{E}) that the energy of a flare depends 
on the height assumption, and so do the parameters $f_0$ and $\delta$ in 
Eq.~(\ref{f(E)}). In the first and second part of this section we study the
scaling relations and conversions between flare energy distributions for
different assumptions for $h$ as well as for time variations of individual
pixels. In the third part of this Section we will apply these relations to numerical values from
observations.     
 
\subsection{Distribution in flare area and per pixel}  \label{area_and_pixel}  
 
Because both the emission measure increase and the area of a flare are 
observable quantities, we write   
\begin{equation} \label{M(A)}  
  \Delta{\cal M}(A) = m A^\mu.   
\end{equation}  
In terms of the above variables we have 
\begin{eqnarray}
  \mu &=& \frac{\gamma}{\alpha} \label{cond3} \\ 
    m &=& c\cdot a^{-\mu}.         \label{cond4}
\end{eqnarray}
There are four equations (Eqs.~\ref{cond1}, \ref{cond2}, \ref{cond3} and
\ref{cond4}) and six variables ($a$, $b$, $c$, $\alpha$, $\beta$, $\gamma$), thus the
system is still not fully determined. This is because we have the
freedom to make an assumption regarding the height dependence. A model for the flare height 
fixes $b$ and $\beta$, and the system of equations 
can be solved. There is one more difficulty, however, and that is to determine
$\mu$ and $m$. 
As Eq.~(\ref{M(A)}) is a direct relation between variables 
of individual events and not a distribution, the scatter is large, which is
obvious in Fig.~\ref{mfaf}. 
There is another way to determine $\mu$ and $m$, using the 
comparison between flare distribution and the energy distribution of
emission measure increases observed in individual pixels. In the following, we
refer to the latter  
as pixel distribution and mark it with the index $p$. 

The pixel distribution function accounts for events in pixels without 
adjacent pixels being grouped together to form one flare. Thus for
flares covering more than one pixel, the pixel 
distribution is the flare distribution times the average flare area measured 
in units of pixel area $A_p$  
\begin{equation} \label{f_p(E_p)} 
  f_p(E_p) dE_p = \frac{A(E)}{A_p} f(E) dE, 
\end{equation}  
where the thermal energy of a pixel is $E_p=3k_BT\sqrt{\Delta{\cal M}_p A_p
  h}$. The average emission measure increase in a pixel is simply $\Delta{\cal
  M}_p=A_p/A\cdot\Delta{\cal M}$. It follows that   
\begin{equation} \label{Ep}  
  E_p = \frac{A_p }{A(E)} E . 
\end{equation}  
We insert Eq.~(\ref{A(E)}) into Eq.~(\ref{Ep}) and solve 
for $E$ to obtain an average relation between the thermal energy input in 
flares and in pixels   
\begin{equation} \label{E(E_p)}  
  E = \left( \frac{a}{A_p}E_p \right)^\frac{1}{1-\alpha}.  
\end{equation}  
Taking the derivative  
\begin{equation} \label{dE/dE_p}  
  \frac{dE}{dE_p} = \frac{1}{1-\alpha} \left( \frac{a}{A_p} 
  \right)^\frac{1}{1-\alpha} E_p^\frac{\alpha}{1-\alpha},   
\end{equation}  
rewriting Eq.~(\ref{f_p(E_p)}) as 
\begin{equation} \label{f_p(E_p)2}  
  f_p(E_p) dE_p = \frac{A(E[E_p])}{A_p} f(E[E_p]) \frac{dE}{dE_p} dE_p  
\end{equation}  
and inserting Eqs.~(\ref{f(E)}), (\ref{A(E)}), (\ref{E(E_p)}) and
(\ref{dE/dE_p}) into Eq.~(\ref{f_p(E_p)2}), leads to the final result for the
relation between the flare distribution and the pixel distribution  
\begin{eqnarray}  \label{pix_conv} 
  f_p(E_p) & = & \frac{1}{1-\alpha} \left( \frac{a}{A_p} 
  \right)^\frac{2-\delta}{1-\alpha} f_0 
  E_p^{-\frac{\delta-2\alpha}{1-\alpha}} \\
           & =: & f_{p0} E_p^{-\delta_p}. \nonumber
\end{eqnarray}  
The exponent $\alpha$ and the factor $a$ then are
\begin{eqnarray}
  \alpha & = & \frac{\delta_p-\delta}{\delta_p-2} \\
       a & = & A_p\left 
  ( \frac{f_{p0}}{f_0}\cdot\frac{\delta-2}{\delta_p-2}\right)
       ^\frac{-1}{\delta_p-2}.  
\end{eqnarray}
Equation (\ref{cond1}) together with the height assumption used to derive the
thermal energy then yields $\gamma$, and $\mu$ 
follows immediately     
\begin{equation} \label{mu} 
  \mu = \frac{\gamma}{\alpha} =
  \frac{\delta_p+\delta-4-\beta(\delta_p-2)}{\delta_p-\delta}.   
\end{equation}  
Similarly, one derives 
\begin{equation} \label{m} 
  m = \frac{a^{-(1+\mu)}}{b (3k_BT)^2}\ \ .  
\end{equation}  

\subsection{Scaling relationships}  
 
The primary free model parameter is the height of the flare 
(Eq.~\ref{h(E)}), i.~e.~the parameters $b$ and $\beta$. They determine the
thermal energy by Eq.~(\ref{E}) when $T$, $\Delta{\cal M}$ and $A$ are
observed. In the following we derive the conversion between the flare
frequency distributions of different height dependencies.  

Let the new flare distribution be 
$f^\prime(E^\prime)$, where $E^\prime=3 k_B T \sqrt{\Delta{\cal M} A 
  h^\prime}$. Then the conversion from the old flare distribution 
(Eq.~\ref{f(E)}) to the new one is given by   
\begin{equation} \label{f'(E')}  
  f^\prime(E^\prime) dE^\prime = f(E[E^\prime])\frac{dE}{dE^\prime} dE^\prime.   
\end{equation}  
It is seen immediately that the identity $E/\sqrt{h}=E^\prime/\sqrt{h^\prime}$ 
holds. Inserting Eq.~(\ref{h(E)}), we obtain from it the condition   
\begin{equation} \label{bE}  
  b^{-\frac{1}{2}} E^{1-\frac{\beta}{2}} = (b^\prime)^{-\frac{1}{2}} (E^\prime)^{1-\frac{\beta^\prime}{2}}.  
\end{equation}  
Solving for $E$ leads to  
\begin{equation} \label{E(E')}  
  E(E^\prime) = \left( \frac{b}{b^\prime} \right)^\frac{1}{2-\beta} E^{\prime\frac{2-\beta^\prime}{2-\beta}}.  
\end{equation}  
The derivative is  
\begin{equation} \label{dE/dE'}  
  \frac{dE}{dE^\prime} = \frac{2-\beta^\prime}{2-\beta} \left( \frac{b}{b^\prime} \right)^\frac{1}{2-\beta} E^{\prime\frac{\beta-\beta^\prime}{2-\beta}}.  
\end{equation}  
Inserting now Eqs.~(\ref{f(E)}), (\ref{E(E')}) and (\ref{dE/dE'}) into 
Eq.~(\ref{f'(E')}) gives the new flare distribution   
\begin{eqnarray}  \label{E_conv} 
  f^\prime(E^\prime) & = & \frac{2-\beta^\prime}{2-\beta} \left
  ( \frac{b}{b^\prime} \right)^\frac{1-\delta}{2-\beta} f_0
  E^{\prime-\frac{\delta(2-\beta^\prime)-\beta+\beta^\prime}{2-\beta}} \\
  & =: & f_0^\prime E^{\prime-\delta^\prime}\ , \nonumber
\end{eqnarray}  
defining the new power law exponent $\delta^\prime$ and $f_0^\prime$. 

\subsection{Numerical values}  

The observed quantities are the emission measure, the flare area and the
temperature. \citet{benkru01} find an average temperature of $1.46\cdot
10^6\,{\rm K}$ for the 23 largest nanoflares that occurred in a quiet
region within one hour. The temperature does not depend on the flare energy,
and its distribution is narrow. The highest observed value was reported to be  $T=1.63\cdot 10^6\,{\rm K}$. We use the former value for our models and will discuss it more in
Sect.~\ref{discussion}. The pixel area is given by $A_p=1900\,{\rm km}\times
1900\,{\rm km}$.  

Making an
assumption on the flare height, the flare energy, flare distribution, pixel
energy and pixel distribution can be reduced, and thus also the parameters $m$
and $\mu$ defined in Eq.~(\ref{M(A)}) according to
Sect.~\ref{area_and_pixel}. Once we know $m$ and $\mu$, we can solve for the
parameters $a$, $b$, $c$ and $\alpha$, $\beta$, $\gamma$. Also, we can
calculate a new flare distribution from Eq.~(\ref{E_conv}) and a new pixel
distribution from Eq.~(\ref{pix_conv}) for a different flare height. This 
shows the effect of the model assumption on the distribution and characteristic energies. Finally, 
the biggest interest is in the parameters for a model with flare height that scales 
with flare size.

Table \ref{table} displays some parameters of interest. The values for $f_0$ and
$\delta$ for $h=5000$\,km are from \citet{kruben98} and $f_{p0}$ and 
 $\delta_p$ for $h=500$\,km are from Krucker (private communication). All the
 other 
values from  $f_0$ to $\gamma$ are calculated using the methods of the previous
sections. 

\begin{figure}[t]  
  \centerline{\psfig{figure=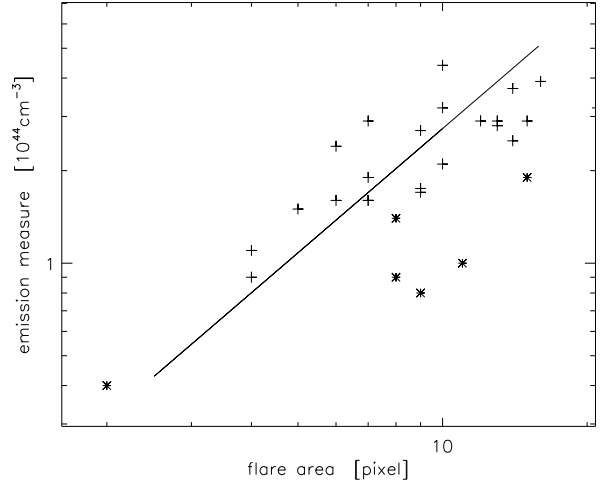,height=7cm}}  
  \caption{{\em Emission measure per flare versus flare area. The crosses are 
      the largest flares observed in a large field of view and the asterisks 
      refer to the largest events in a small field of view 
      \citep{kruben00}. The solid line represents the relation $\Delta{\cal M}
      = m A^\mu$ (Eq.~\ref{M(A)}), using the parameters calculated from
      Eqs.~(\ref{mu}) and (\ref{m}) for $\beta=0$.}\label{mfaf}}     
\end{figure}

First, $f_0(500\,{\rm km})$ and $\delta(500\,{\rm km})$ have been calculated from
Eq.~(\ref{E_conv}), then $m$ and $\mu$ from Eqs.~(\ref{m}) and
(\ref{mu}). The relation of Eq.~(\ref{M(A)}) is displayed in Fig.~\ref{mfaf}
(solid line).
This figure also contains the largest 22 events 
reported by \citet{kruben00} from a large field of view, thus the emission
measures are relatively large (their Table I, shown here by $+$), and the
largest 6 
events from a smaller set (their Table II, shown by $\ast$). The calculated
relation between emission measure and flare area matches well that observed
for individual events. Knowing the parameters $m=7\cdot 10^{20}$ and $\mu
=1.34$,  
which are independent of the height model, all the other
parameters can 
then immediately be calculated for any chosen height dependence.

Rather large nanoflares are seen to cover an area of
approximately rectangular shape with a length to width ratio $q=\l/w$ of about
10. Thus 
\begin{equation}
 \label{Alw} 
  A = l\cdot w \approx l^2/10 = 10w^2.
\end{equation}
As ordinary flares are believed to be generated in loop-shaped magnetic field 
structures, nanoflares may be expected to originate in similar 
geometries. Thus, we assume the nanoflares to occur in torus-shaped loops with
a loop-thickness equal to the flare width. In our approximation this
loop-thickness shall just be the unknown effective height, thus $h=w=\sqrt{(A/10)}$ and
$\beta=\alpha/2$ and $b=\sqrt{(a/10)}$. 

The largest single-peaked flare observed was 
derived to have an energy of $E_{max}(5000\,{\rm km})$ = $2\cdot 10^{26}$\,erg 
\citep[Table~I]{kruben00}. The minimum energy can be derived by equating the
total radiation output $P_{tot}=4.5\cdot 10^{5}\,{\rm erg}^{-1}\,{\rm
  cm}^{-2}\,{\rm s}^{-1}$ with the integrated flare energy input. This value
was derived by \citet{kruben98} from the total emission measure observed in
coronal EUV lines, averaged over the whole field of view, and includes all
radiation losses in the continuum and the lines from UV to X-rays. Solving     
\begin{equation}  \label{Ptot}
   P_{tot} = \int_{E_{min}}^{E_{max}} E\cdot f(E) dE   
\end{equation}  
one obtains for $\delta >2$
\begin{equation}  
  E_{min} \approx \left( \frac{\delta-2}{f_0}
  P_{tot}\right)^{-\frac{1}{\delta-2}}. 
\end{equation} 
Here we have used the fact that for our parameter range the upper boundary term
of Eq.~(\ref{Ptot}) is vanishingly small.  
The length of the smallest flares is given by Eqs.~(\ref{A(E)}) and (\ref{Alw})
\begin{equation}
  l_{min} = \left( 10\cdot a\cdot E_{min}^\alpha\right)^{1/2}.
\end{equation}
If the flare size is equal to the area of a pixel, Eq.~(\ref{Ep}) shows that the 
flare energy is equal to the energy in the pixel increase. At this energy,
$E_{res}$, the flare energy distribution will have a low-energy roll-over due
to sub-resolution events. The energy $E_{res}$ is obtained by solving
Eq.~(\ref{E(E_p)}) for $E=E_p\equiv E_{res}$. It is consistent with the cutoff
energy of $9\cdot 10^{24}$\,erg observed by \citet[Fig.~2, assuming $h$ = 5000
km]{kruben98}.      
\begin{table}[t]
\caption{{\sl Observed and derived parameters of nanoflare energy distributions 
for different height models given by $h$, $b$ and $\beta$. The temperature
used was $T=1.56\cdot 10^6{\rm K}$. \label{table}}}   
\begin{tabular}{lrrr}  
$h$        & 5000\,km            & 500\,km            & $\sqrt{(A/10)}$       \\ 
\hline 
$f_0$      & $10^{19.2}$         & $10^{18.4}$        & $3.38\cdot 10^{11}$ \\ 
$\delta$   & 2.59                & 2.59               & 2.31                \\  
$f_{p0}$   & $10^{106.12}$       & $10^{103.6}$       & $1.30\cdot 10^{30}$ \\  
$\delta_p$ & 6.04                & 6.04               & 3.05                \\ 
\hline 
$a$        & $1.8\cdot 10^{-5}$  & $4.7\cdot 10^{-5}$ & 0.209               \\ 
$b$        & $5\cdot 10^8$       & $5\cdot 10^7$      & 0.145               \\ 
$c$        & $2.7\cdot 10^{14}$  & $1.0\cdot 10^{15}$ & $7.91\cdot 10^{19}$ \\ 
$\alpha$   & 0.854               & 0.854              & 0.704               \\ 
$\beta$    & 0                   & 0                  & 0.352               \\ 
$\gamma$   & 1.146               & 1.146              & 0.944               \\ 
\hline 
$E_{min}$  & $2.2\cdot 10^{23}$\,erg & $9.8\cdot 10^{21}$\,erg & $3.7\cdot 
10^{20}$\,erg  \\ 
$E_{res}$  & $9.0\cdot 10^{24}$\,erg & $2.8\cdot 10^{24}$\,erg & $3.1\cdot
10^{24}$\,erg  \\ 
$E_{max}$  & $2.0\cdot 10^{26}$\,erg & $6.3\cdot 10^{25}$\,erg & $1.3\cdot
10^{26}$\,erg \\  
$l_{min}$  & 1230\,km             & 530\,km             & 250\,km            
\end{tabular} 
\end{table} 

\begin{figure}[t] 
   \centerline{\psfig{figure=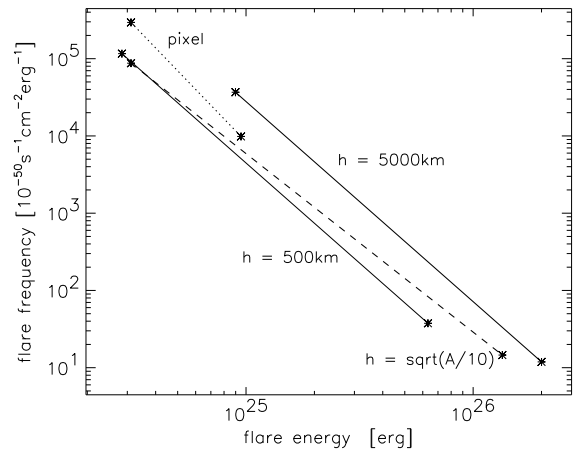,height=6.5cm}} 
   \caption{{\sl Various flare distributions and the pixel distribution for
       $h=\sqrt{A/10}$ are 
       shown for different height assumptions. The bold line labeled with 
       $h$=5000\,km corresponds to the flare energy distribution observed by 
       \citet{kruben98}. The other distributions are derived using 
       Eqs.~(\ref{pix_conv}) and (\ref{E_conv}). The asterisks denote
       $E_{res}$ and $E_{max}$.}}  
   \label{freqdistr} 
\end{figure} 
In Fig.~\ref{freqdistr} the different flare distributions as well as the pixel
distribution for $h=\sqrt{(A/10)}$ are shown for comparison. The distributions
are displayed for energies between $E_{res}$ and $E_{max}$ as they were
observed. The pixel distribution only makes sense for energies
higher than the resolution energy, because only these flares cover one or more 
pixels. This can be illustrated by the following consideration: Let us
assume a flare with energy smaller than $E_{res}$, thus the flare area is
smaller than the pixel area. According to Eq.~(\ref{f_p(E_p)}), the number of
pixel events in the energy interval $dE_p$ would be smaller than the number of
flare-events in the corresponding energy interval $dE$. This contadicts the assumption that, when any such flare occurs, it has to appear
in exactly one pixel. We neglect here the effect of subresolution flares spreading over pixel boundaries. Thus, for flares with area/energy smaller than the resolution, the pixel distribution is equal to the flare distribution and not
given by Eq.~(\ref{f_p(E_p)}), which in that case makes no sense. This can
also be visualized by the definition for the flare distribution, which is the
number of flares divided by total observed area, time and energy
interval. Obviously, for flares within the observed area this
fraction is constant. Reducing the area to the size of a pixel doesn't change
the flare frequency distribution for flares with still smaller size.

 

\section{Simulation} \label{simulation}

In this section we simulate numerically the emission measure variations in
individual pixels. The goal of the simulation is to find the conditions for
making the simulated pixels resemble the observed emission measure in form
and background level, and to make the averaged simulated power spectrum match
the observed one in slope and absolute value. 

It is assumed that the total emission measure in a pixel at a certain time is given by the sum of
the emission measures of all flares brightening at that time. The flares are
described by their energy, which is distributed according to the observed
pixel distribution for flare energies in the observed range $E_{res}$ to
$E_{max}$ and flare distribution for the extrapolated sub-resolution events
from $E_{min}$ to $E_{res}$. The emission measure increase is given by
Eq.~(\ref{M(E)}). Because this increase was defined as the difference between a
maximum and the preceeding minimum of a brightening, it is about equal to the
maximum value of the time dependent emission measure of a flare. The emission
measure of a flare increases up to this value and then decreases again
according to the flare's time profile.

In the simulation, the flares are randomly distributed in time, according to
the flare and pixel distribution. Each flare is defined by its maximum value and the time profile. For
every observation, the total emission measure then is the sum of the
respective flares' emission measures at that time. To make adequate comparison
with the observations, we choose the total observing time $t_{obs}$ and the
equally distributed number of observing points $N_{obs}$ in the same way as in
the observations by \citet{benkru98}. Of the total time dependent emission 
measure, the power
spectrum is taken numerically. This process, simulating a pixel and taking the
power spectrum, is repeated many times and the resulting power spectra are 
averaged. The results, the single pixel emission measure level and
variations and the average power spectrum, are then compared with the
observations.

\subsection{Time profile} \label{time_profile}

The power spectrum of the emission measure variations in time per pixel depends
on the time profile of individual events. Let ${\cal M}$ be the emission
measure of one flare. The assumed self-similarity of flares requires that
${\cal M}$ must depend on the average only on energy in amplitude as well as in duration. The
temporal evolution of a flare is given by the rate at which plasma is heated
to coronal temperatures, expands and cools. We will assume that the duration
is proportional to $E^s$. The average emission measure of flares with
energy $E$ is therefore 
\begin{equation} \label{M(t)}
  {\cal M}(E,t) = c E^\gamma g(E^{-s}t), 
\end{equation}
where $g$ is the time profile. Note that the maximum value of the emission
measure of one large flare is assumed to be the observed emission measure
increase. Therefore the time profile $g(x) = g(E^{-s}t)$ must be chosen in
such a way, that $max(g) =1$. The emission measure increase $\Delta{\cal M} =
max({\cal M}) = c E^\gamma$ then follows the relationship Eq.~(\ref{M(E)}).  

Let $\tau(E)$ be the characteristic duration of a flare with thermal energy
$E$. We then have
\begin{equation} \label{tau1}
  \tau(E) \propto \frac{E}{\dot{E}} \propto E^s. 
\end{equation}
The emission measure ${\cal M}$ is proportional to the observed energy loss
rate from radiation. The corona may also lose energy from conduction at the
foot-points of the loop with a rate $Q$.  Then $\dot{E}\propto {\cal
  M}+Q$. Assuming that $Q$ follows the same power law as ${\cal M}$ or is much
smaller and therefore can be neglected, we obtain 
\begin{equation} \label{tau}
  \tau(E) \propto \frac{E}{{\Delta\cal M}(E)}. 
\end{equation} 
Inserting Eq.~(\ref{M(E)}) and comparing with Eq.~(\ref{tau1}), we immediately
obtain $s=1-\gamma$. The simulation results bear out this assumption.

\subsection{Density} \label{density}

In the derivation of the equation for the thermal energy of a flare
(Eq.~\ref{E}) it was assumed that the density $n(E)$ is approximately
constant in the flare volume. Thus, 
\begin{equation}
  \Delta{\cal M} := \int_V n^2 dV \approx n^2 V\ ,
\end{equation}
where $V=A\cdot h=abE^{\alpha+\beta}$. Solving for $n$ and using
Eq.~(\ref{cond1}), we get 
\begin{equation}
  n = \left( \frac{c}{ab}\right) E^{-s}.
\end{equation}
For $\gamma < 1$ ($s>0$), the density decreases with energy as the duration
increases. A physical explanation may be that the longer life time of large
flares is partially a result of lower density. Note that the effect is small
as $s$ is close to zero in all models. A similar result has been derived by
\citet{atnetal00}. 

\subsection{Results} \label{results}

In the simulation we use the area dependent height model $h=\sqrt{A/10}$
with the parameters derived from observations (Table \ref{table}, $3^{\rm rd}$
column).
We just remark here that this is not the only height assumption which yields
reasonably good results. In particular, the observations are also reproducible
for a model with constant height with only minor adjustments in other
parameters. 
Note that the factor $c$, which is directly proportional to the emission
measure (Eq.~\ref{M(E)}) and therefore proportional to the square root of
the power spectrum, is inversely proportional to the temperature squared. Thus,
the absolute value of the power spectrum is proportional to the minus fourth power of the temperature. For fine tuning the absolute level of the power spectrum we will
adjust the temperature in the simulation. This is equivalent to saying that the
effective thermal energy is not given by Eq.~(\ref{E}), but rather by
$E=\epsilon 3k_BT_{obs}\sqrt{\Delta{\cal M} A h}$, with $T_{obs}$ the observed
temperature and $\epsilon$ a correction variable for the true thermal
energy. Thus, the temperature we use is $T=\epsilon T_{obs}$ with $T_{obs}$ = 1.46$\cdot 10^6$K.

The following time profile is used
\begin{equation} \label{timeprofile}
  g(x) = 
  \left\{ \begin{array}{rl}
            {\rm exp}(\xi x), & x<0 \\
            {\rm exp}(-\eta x), & x\ge 0
          \end{array}
  \right.
\end{equation}
with $x=\frac{t-t_i}{\tau}$, $t_i$ the peaking time of the $i$-th flare with
energy $E$ and duration $\tau=\tau(E)=(E/E_0)^s\tau_0$, $\tau_0=1000\,{\rm s}$
the characteristic time of a flare with energy $E_0=5.8\cdot 10^{25}\,{\rm
  erg}$, $s=1-\gamma=0.0556$ and $\xi$ and $\eta$ positive free
parameters. For the presented results we have used $\xi=20$ and $\eta=0.8$.
The emission measure
per pixel is defined as $EM:={\cal M}/A_p$ and is of order of $10^{27}\,{\rm
  cm}^{-5}$. The pixel area is $A_p=3.61\cdot 10^{16}\,{\rm cm}^2$, and the
total observation time is $t_{obs}=N_{obs}\Delta t$ with $N_{obs}=21$ the number
of observations and $\Delta t=121.9\,{\rm s}$ the time step. 

\begin{figure}[t]
   \centerline{\psfig{figure=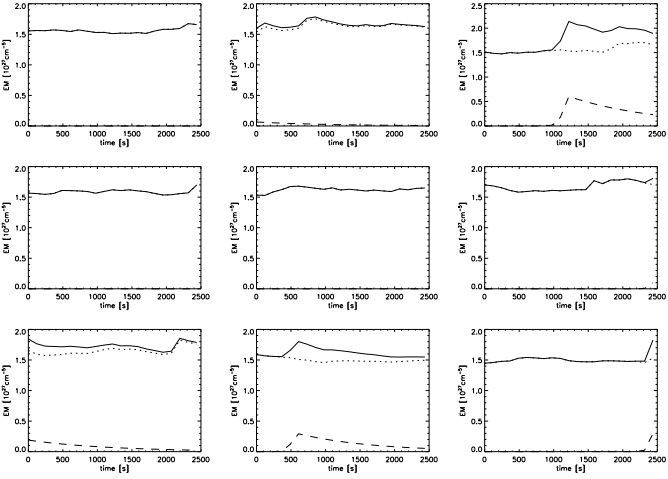,height=6.5cm}}
   \caption{{\sl Emission measure per pixel (solid curve) displayed for
       random pixels. The dotted curve is background from sub-resolution
       flares. The 
       spatially resolved nanoflares are shown by a 
       dashed curve.} 
   \label{pixel}}
\end{figure}

In Fig.~\ref{pixel} some simulated pixels are displayed. The dotted curve
is the summed emission measure over the flares in the sub-resolution
regime. We see that they are responsible for the background level. The dashed
curve occurring occasionally is a resolved nanoflare. The solid curve is the sum
of the two: the total emission measure per pixel.

\begin{figure}[t]
   \centerline{\psfig{figure=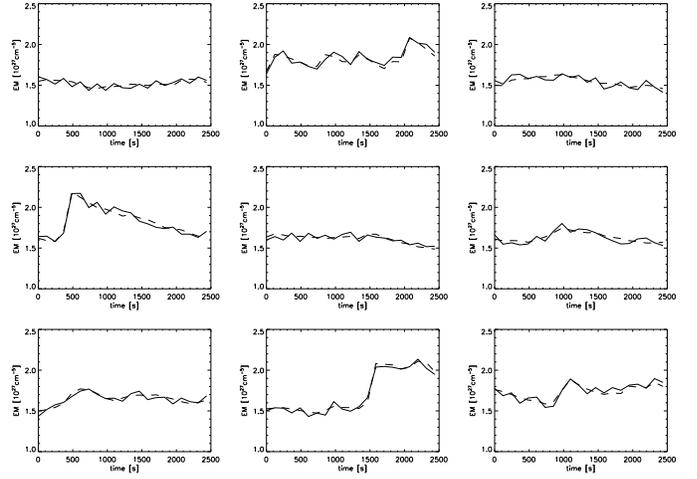,height=6.5cm}}
   \caption{{\sl Noise added emission measure per pixel (solid curve) displayed
       for random pixels (different from Fig.~\ref{pixel}). The dashed curve is 
       the emission measure without noise.} 
   \label{pixelvar}}
\end{figure}

To simulate the effect of noise, the total emission measure at an observed
time is distributed 
randomly within the emission measure noise interval $dEM$. 
Figure \ref{pixelvar} shows the effect of noise for the same model parameters. 

\begin{figure}[t]
   \centerline{\psfig{figure=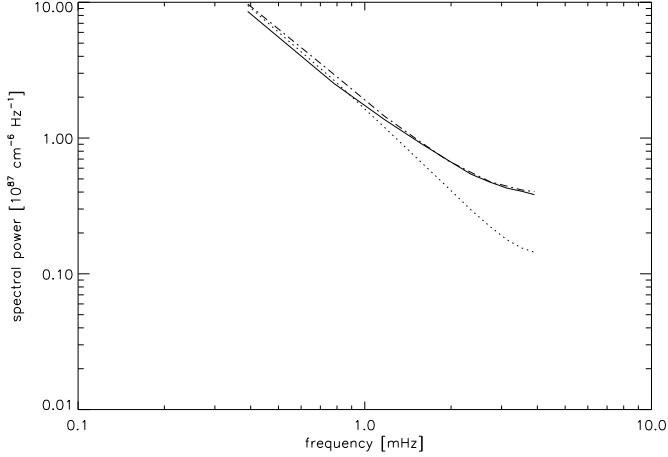,height=6.5cm}}
   \caption{{\sl Simulated power spectrum averaged over 10\,000 pixels with
       $\xi=20$, $\eta=0.8$. The
       dotted curve is the power spectrum without noise and the dashed-dotted
       curve the one with noise, the latter is to be compared with the observed power
       spectrum (solid curve).} 
   \label{power}}
\end{figure}

Figure \ref{power} shows the averaged power spectrum of pixels, simulated with
the same parameters as in Figs. \ref{pixel} and \ref{pixelvar}. The solid
curve was observed by \citet{benkru98} averaged over all pixels, and here
corrected in the ordinate.  The slope of the simulated power spectrum (dotted
curve) at low frequencies is mainly given by the choice of the free parameters
$\xi$ and $\eta$ (see also Sect.~\ref{discussion}). The noise increases the
power at all frequencies (dashed-dotted curve). At low frequencies the slope
is not changed, but at high frequencies the curve is considerably bent upward,
consistent with observations. The absolute values of the power spectrum depend
strongly on the temperature as discussed in the beginning of this
subsection. The best fit is found to be $T\approx 1.56\cdot 10^6\,{\rm K}$, thus  $\epsilon\approx 1.068$.

The averaged power spectrum was found to be highly 
sensitive to the time profile of a flare, which is poorly known from direct
observation. 

\section{Analytic power spectrum} \label{powerspectrum}

In this section the power spectrum of a given nanoflare model distribution is 
calculated analytically. 

\begin{figure}[t]
   \centerline{\psfig{figure=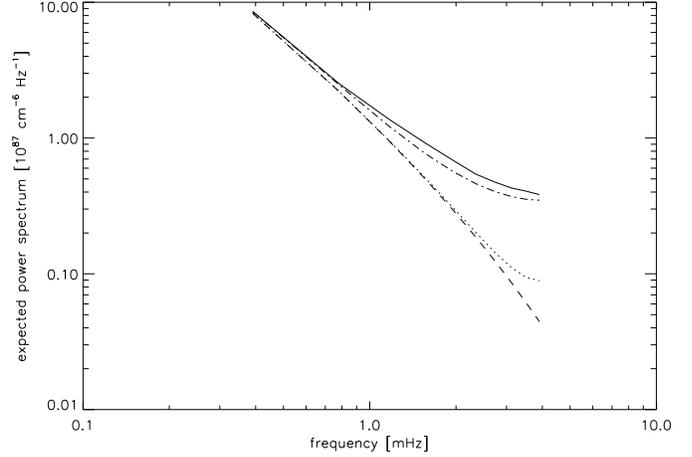,height=6.5cm}}
   \caption{{\sl Analytically calculated power spectrum with $\xi=20$,
       $\eta=0.8$. The dashed curve shows the 
       true power spectrum, the dotted curve the power spectrum with aliasing
       and the dashed-dotted curve the power spectrum with aliasing and
       noise. The solid curve is the observed power spectrum.} 
   \label{spectrum}}
\end{figure}

Let the expected value of the averaged power spectrum at frequency $\omega$ be 
${\cal  E}(|\widehat{{\cal M}}_{tot}(\omega)|^2)$, where $\widehat{{\cal M}}_{tot}$
denotes the Fourier transform of the emission measure of a pixel. If the
flares follow a Poissonian distribution, are independent of each other and
identical at the same energy, it can be shown that the total power spectrum is
composed of a flare component and a noise component (see Appendix
\ref{appendix})
\begin{equation}
  {\cal E}(|\widehat{{\cal M}}_{tot}(\omega)|^2) = 
  {\cal E}(|\widehat{{\cal M}}_{fl}(\omega)|^2) + 
  {\cal E}(|\widehat{{\cal M}}_{noise}(\omega)|^2)
\end{equation}
with
\begin{equation}
  {\cal E}(|\widehat{{\cal M}}_{fl}(\omega)|^2) = A_p
  \int_{E_{min}}^{E_{max}} 
                f(E) |\widehat{{\cal M}}(E,\omega)|^2 dE
\end{equation}
and
\begin{equation}
  {\cal E}(|\widehat{{\cal M}}_{noise}(\omega)|^2) = 
  \frac{\Delta t}{12}(d{\cal M})^2,
\end{equation}
$d{\cal M}$ being the noise level.
Note that in the simulation as well as in the observation the numerical Fourier
transform is normalized by division of the total observing time $t_{obs}$, so
here we use the same normalization. Again we consider pixels. Thus the pixel
distribution has to be taken for large events, and the integral splits into
two parts, 
\begin{eqnarray}
  {\cal E}(|\widehat{{\cal M}}_{fl}(\omega)|^2) &=& 
   A_p \left[ 
  \int_{E_{min}}^{E_{res}} f(E) |\widehat{{\cal M}}(E,\omega)|^2 dE + \right.
  \nonumber\\
  &+&\left.\int_{E_{res}}^{E_{max}} f_p(E_p) |\widehat{\cal M}_p(E,\omega)|^2
    dE_p\right].
\end{eqnarray}
With $f_pdE_p=A/A_p\cdot fdE$ and ${\cal M}_p=A_p/A\cdot {\cal M}$ we get
\begin{eqnarray}
  {\cal E}(|\widehat{{\cal M}}_{fl}(\omega)|^2) &=& 
   A_p \left[ 
  \int_{E_{min}}^{E_{res}} f(E) |\widehat{{\cal M}}(E,\omega)|^2 dE + \right.
  \nonumber\\
  &+&\left.\int_{E_{res}}^{E_{max}} \frac{A_p}{A(E)}
    f(E) |\widehat{{\cal M}}(E,\omega)|^2 dE\right]. \label{expect_value}
\end{eqnarray}
The Fourier transform of a single flare is given by
\begin{eqnarray}
  \widehat{{\cal M}}(E,\omega) &=& 
  c E^\gamma \widehat{g(t/\tau)} \\
  &=& c E^\gamma \tau \widehat{g}(\tau\omega).
\end{eqnarray}
With $g(x)$ from Eq.~(\ref{timeprofile}) we have
\begin{equation}
  \widehat{g}(k) = \frac{\xi+\eta}{(\xi-{\rm i} k)(\eta+{\rm i} k)}
\end{equation}
with $k=\tau\omega$ and
\begin{equation}
  |\widehat{{\cal M}}(E,\omega)|^2 = 
  c^2 E^{2\gamma} \tau(E)^2
  \frac{(\xi+\eta)^2}{(\xi^2+(\tau\omega)^2)(\eta^2+(\tau\omega)^2)}.
\end{equation}
Using $\omega=2\pi\nu$ and inserting all into Eq.~(\ref{expect_value}), we
obtain  \pagebreak
\begin{equation}\label{exp_power}
  {\cal E}(|\widehat{{\cal M}}_{fl}(\nu)|^2) = 
    A_p c^2 \frac{\tau_0^2}{E_0^{2s}} f_0 (\xi+\eta)^2 \cdot 
\end{equation} \nopagebreak
\begin{eqnarray}
   \cdot\left[\int\limits_{E_{min}}^{E_{res}}
    \frac{E^{2-\delta}dE}
    {(\xi^2+(2\pi\tau_0\left(\frac{E}{E_0}\right)^s\nu)^2)
      (\eta^2+(2\pi\tau_0\left(\frac{E}{E_0}\right)^s\nu)^2)} 
    +\right.\nonumber\\
  \frac{A_p}{a}\left.\int\limits_{E_{res}}^{E_{max}}
    \frac{E^{2-\delta-\alpha}dE}
    {(\xi^2+(2\pi\tau_0\left(\frac{E}{E_0}\right)^s\nu)^2)
      (\eta^2+(2\pi\tau_0\left(\frac{E}{E_0}\right)^s\nu)^2)} 
    \right] \nonumber
\end{eqnarray}
which can be computed numerically. 
The solution is shown in Fig.~\ref{spectrum}, dashed curve, with all parameters
the same as in Sect.~\ref{results}, in particular $\xi=20$
and $\eta=0.8$. In the dotted curve,
aliasing is additionally taken into account. The noise term ${\cal
  E}(|\widehat{{\cal M}}_{noise}(\omega)|^2)$ is added to the aliased power
spectrum (dashed-dotted 
curve) with $d{\cal M}=A_p\cdot dEM$. The solid
curve shows the observed power spectrum. 

\section{Discussion} \label{discussion}

It is clear from the previous two sections that a model exists that 
can explain the emission measure and its fluctuations observed in the quiet
solar corona. In this section we address the question of how sensitive the
results are to the choice of parameters. Assumptions were made on the height
dependence of the thermal energy, extrapolation of the flare distribution to
energies several orders of magnitude smaller than the observed range and the
shape and energy dependence of the time profile of single pixels. Also, an
assumption 
used throughout this model is the self-similarity of flares at different
energies in the average, which is further supported by having found a set of
parameters 
that explains all the available observations. 
In the following considerations, we will focus on a model with $h=\sqrt{A/10}$,
which seems to be a plausible choice. However, it should be mentioned that
this model is not unique in meeting the above requirements. In particular,
models with a constant height of 500\,km or 5000\,km, respectively, also yield
reasonably good results. The remaining free parameters for fitting , $T, \xi$, and $\eta$, however remain practically the same.

The three main observational constraints mentioned in the introduction are 
the typical emission measure fluctuation 
of a pixel, its absolute value in emission measure and the shape and value of
the power spectrum. As seen in Fig.~\ref{pixel}, the sub-resolution flares
contribute most to the background level of the emission measure. If we raise the
minimum energy or, equivalently, reduce the total input power, the required
number of 
small flares decreases and the background emission level drops drastically. To
explain the observed level in emission measure we need to extrapolate the
energies to $E_{min}$. The background level of the preferred model yields values 
at the low end of the observed emission measure in the quiet corona. A
constant emission measure background  
(e.~g.~produced by another heating process) could be added without changing the 
power spectrum.

The nanoflare time profile plays an important role in all three aspects. The
{\sl background level} in emission measure depends on the duration of a flare. A
slow increase and decrease in the time profile makes a flare contribute longer
to the background, whose level then rises. In this case the fluctuations
become smaller. For a fast increase and decrease the background level drops
and the fluctuations become more peaked and accentuated. 

The rate of change has a strong effect on the 
slope in the {\sl power spectrum}.   
\begin{figure}[t]
   \centerline{\psfig{figure=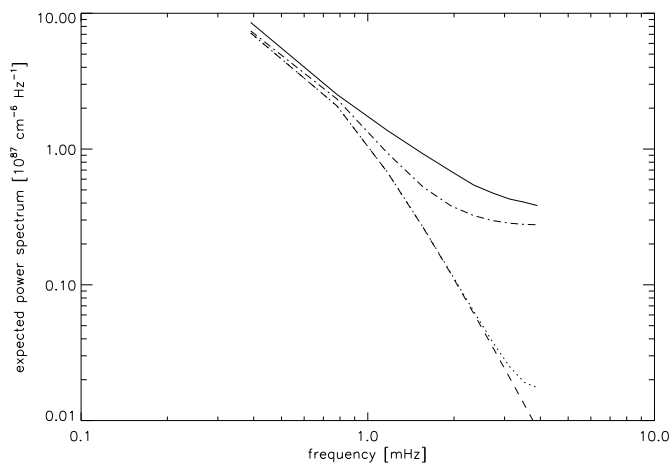,height=6.5cm}}
   \caption{{\sl Analytic power spectrum for a symmetric nanoflare time profile
       with $\xi=\eta=3.41$. The dashed curve displays the basic power spectrum,
       the dotted curve the power spectrum with aliasing and the dashed-dotted
       curve the one with noise and with aliasing. This final calculated curve 
       drops drastically at higher frequencies, not even noise can lift it
       enough to match the observed spectrum (solid curve) at high frequencies.} 
   \label{sym}}
\end{figure}
For the time profile given by Eq.~(\ref{timeprofile}), the slope of the power
spectrum is mainly determined by the free parameters $\xi$ and $\eta$. A slope
with the observed value $\kappa=-1.76$ between 
the lowest two frequency points can be obtained by a wide parameter range,
e.~g.~$(\xi,\eta)=\{(3.41,3.41), (10,1.2), (15,1.0), (20,0.8),
(50,0.7)\}$. For a symmetric time profile, i.~e.~$\xi=\eta=3.41$, the power at high
frequencies drops so drastically, that a bump appears at medium frequencies
that cannot be flattened out by noise (Fig.~\ref{sym}). The more asymmetric
the time 
profile, the straighter is the slope of the power spectrum and
the more it resembles the observed one. That the time profile should be so
asymmetric 
is surprising, although it is confirmed immediately by
Eq.~(\ref{exp_power}) for the time profile used. If
$\xi$ is large compared to the $\nu$ dependence, one bracket-term in the
denominator of the integral is almost negligible. If, additionally, $\eta$ is
small, the power spectrum is approximately proportional to $\nu^{-2}$. Added noise
reduces the steepness of the slope further which then matches the observed
$\kappa$. The small discrepancies in the level of the simulated and expected
power spectrum come from its sensitive dependence on the parameters $a$,
$b$ and $c$ and numerical errors in the last digits thereof. A small
dependence of $\kappa$ also comes from the maximum energy $E_{max}$. The
larger the maximum energy, the steeper the slope.


\section{Conclusion} \label{conclusion}

Deriving distributions of flare energy and pixel variations from observations
requires some assumptions, in particular a model of the effective
line-of-sight thickness (height). 
We have here developed the analytical tools for the transformation between
different assumptions published in the literature. 
These tools are the basis on which we develop a numerical model that
simulates the time behavior of individual pixels and nanoflares. It is used
here to reproduce the EIT observations of \citet{benkru98} and
\citet{kruben98}. 

We have approached the question of nanoflare heating by searching for a
model of energy input by nanoflares that can reproduce the three constraints
given by the observations of individual pixels: the general appearance of the
emission measure variation in time per pixel, the average level of the
radiation loss, and the average Fourier power spectrum in time.  

We obtain the power spectrum in two different ways, averaging the simulation
and calculating the expected value directly. As the evaluation of the integral
consumes much less calculation time than the simulation, this method allows 
us to test systematically different time profiles. 

The analysis by analytical and numerical studies has shown that there is at
least one such model, assuming self-similarity in energy. Closest to reality may be a model with a height proportional to the square root of the area. 
The model parameters correspond to the last column in Table \ref{table}. 
To reproduce the observed radiation loss, the observed range of nanoflares
(having a lower limit at about $3\cdot10^{24}$\,erg, cf. Fig.~\ref{freqdistr}) needs
to be extrapolated to lower energies by four orders of magnitude (Table
\ref{table}). This energy is far below the model suggested by
\citet{parker83}, although there is no stringent theoretical lower limit of
magnetic energy release. Nevertheless, the extrapolation is hypothetical for
several reasons, in particular the assumption of self-similarity implying a
constant power-law index for the distributions in energy, area and height. Other height assumptions, e.~g.~constant height, also yield reasonable results after slight adjustments for some parameters.

The minimum energy depends on several assumptions. It is larger if the flares
introduce other forms of energy into the corona than just the thermal energy of
evaporated material, such as fluid motions or wave energy. They would heat the
material already existing in the corona and yield a constant background
emission measure. Such a background level would reduce the thermal energy
input requirements by nanoflares and enhance the minimum energy. A test on the
importance of a steady emission measure versus flares could be made by
observations of much higher sensitivity than possible today with EIT and
TRACE. 

A hint of a break-down of our self-similar model may be the small
length of flare loops derived (Table \ref{table}). A possible remedy may be to
assume an 
energy dependence of the length to width ratio $q$. One may expect from a
reconnection scenario that the width and effective height decrease faster with
energy than the length, thus forming the smallest flares in thin, long
loops. There are currently no observations of this ratio over a sufficient
range of flare energies. High resolution observations could clarify this point
in the future.

The fitting model is not unique, particularly in the choice of height
assumption and average time profile of nanoflares. Nevertheless, we found that
the three observational constraints severely limit the range of free
parameters once the model assumptions have been made. In particular, the power
spectrum is found to be very sensitive to the chosen time profile. As the
exact shape of the time profile is not easily observable, this can be used to
test how nanoflares evolve and disappear.


\begin{acknowledgements}
We would like to thank S.~Krucker for additional information and M.~S.~Wheatland for
providing a first simulation program. We also acknowledge K.~W.~Smith for suggestions; E.~Kraev for mathematical
advice; P.~Messmer for answering all questions concerning programming; G.~Paesold for useful comments and M.~G\"udel, M.~Audard and A.~Pauluhn for
general discussions and advice.
The work at ETH Zurich is financially supported by the Swiss National Science
Foundation (grant No. 20-53664.98). 
We make use of previous EIT/SoHO observations and thank the EIT team, in
particular J.-P.~Delaboudini\`ere. EIT was funded by CNES, NASA, and the
Belgian SPPS. 
\end{acknowledgements}

\appendix  
\section{Derivation of the expected value of the power spectrum}  
\label{appendix}    

Consider the observation time interval $t_{obs}$ divided into $N$ subintervals
of equal length through the points $t_0=0, t_1, \ldots, t_N=t_{obs}$, with
$N\gg N_{obs}$. If the subintervals are chosen to be sufficiently small, we
can assume without loss of generality that each flare peaks at one of the
points $t_n$. Further let us divide our energy range into $M$ equal intervals
of length $\Delta E$ through the points $E_0=E_{min}, E_1, \ldots,
E_M=E_{max}$. Let $p_{mn}$ denote the number of flares of energy between $E_m$
and $E_m+\Delta E$ peaking at $t_n$. For any $m$, let $p_{mn}$ be Poissonian,
independent and identically distributed. Their variance is then equal to
their expected value and we have 
\begin{equation} \label{var}
  Var(p_{mn}) = {\cal E}(p_{mn}) = \frac{f(E_m)\Delta E}{\int f(E)dE}
\end{equation}
with $f(E_m)$ the frequency distribution of the flares, normalized to 
\begin{equation}
  \int f(E)dE = \frac{N}{A_p t_{obs}}.
\end{equation}

Let the emission measure of a flare with energy $E_m$ peaking at time $t_n$ be 
\begin{equation}
  {\cal M}_{mn}(t) = {\cal M}(E_m,t-t_n). 
\end{equation}
The total emission measure at time $t$ is then given by
\begin{equation}
  {\cal M}_{fl}(t) = \sum_{m,n}p_{mn}{\cal M}_{mn}(t).
\end{equation}
Thus, we have
\begin{equation}
  \widehat{{\cal M}}_{fl}(\omega) 
   =  \sum_{m,n}p_{mn}\widehat{\cal M}(E_m,\omega)\cdot {\rm e}^{-i\omega
  t_n},
\end{equation}
therefore
\begin{equation} 
  |\widehat{{\cal M}}_{fl}(\omega)|^2  =  
  \widehat{{\cal M}}\cdot \widehat{{\cal M}}^\ast =
\end{equation}
\begin{eqnarray}
   =  \sum_{m_1,m_2\atop n_1,n_2}&&p_{m_1n_1}p_{m_2n_2} \nonumber \\
  &&\widehat{{\cal M}}(E_{m_1},\omega)\widehat{{\cal M}}^\ast(E_{m_2}, \omega)
  {\rm e}^{-i\omega (t_{n_1}-t_{n_2})}.
\end{eqnarray}
As the expected value operator commutes with the sum, we have
\[
  {\cal E}(|\widehat{{\cal M}}_{fl}(\omega)|^2 )  =
\]
\begin{eqnarray}  
  =\sum_{m_1,m_2\atop n_1,n_2}&&{\cal E}(p_{m_1n_1}p_{m_2n_2})\cdot \nonumber \\
  &&\widehat{{\cal M}}(E_{m_1},\omega)\widehat{{\cal M}}^\ast(E_{m_2},\omega)
  {\rm e}^{-i\omega (t_{n_1}-t_{n_2})} \\
  =  \sum_{m_1,m_2\atop n_1,n_2} 
  &&\left[ \delta_{m_1m_2}\delta_{n_1n_2} Var(p_{m_1n_1}) + 
    {\cal E}(p_{m_1n_1}){\cal E}(p_{m_2n_2}) \right]\cdot \nonumber \\
  &&\widehat{{\cal M}}(E_{m_1}, \omega)\widehat{{\cal M}}^\ast(E_{m_2},\omega)
  {\rm e}^{-i\omega (t_{n_1}-t_{n_2})}.
\end{eqnarray}
Inserting Eq.~(\ref{var})
\[
  {\cal E}(|\widehat{{\cal M}}_{fl}(\omega)|^2 )=
\]
\begin{eqnarray}
    = && \frac{A_pt_{obs}}{N}
    \sum_{m,n}f(E_m)
    \widehat{{\cal M}}(E_m,\omega)
    \widehat{{\cal M}}^\ast(E_m,\omega) 
    \Delta E + \nonumber\\
    + && \sum_{m_1,m_2}f(E_{m_1})f(E_{m_2})
    \widehat{{\cal M}}(E_{m_1},\omega)
    \widehat{{\cal M}}^\ast(E_{m_2},\omega)\Delta E^2 \cdot \nonumber \\
    && \frac{(A_pt_{obs})^2}{N^2} 
    \sum_{n_1,n_2}{\rm e}^{-{\rm i}\omega (t_{n_1}-t_{n_2})} 
\end{eqnarray}
and taking the sum over $n_1$, $n_2$ and $n$, respectively, one obtains
\[
  {\cal E}(|\widehat{{\cal M}}_{fl}(\omega)|^2 )=
\]
\begin{eqnarray}
  = && A_pt_{obs}\sum_m  f(E_m) 
  |\widehat{{\cal M}}(E_m,\omega)|^2\Delta E + \nonumber\\
  + && \frac{(A_pt_{obs})^2}{N}\cdot \\
  && \sum_{m_1,m_2}  f(E_{m_1}) f(E_{m_2}) 
  \widehat{{\cal M}}(E_{m_1}, \omega)\widehat{{\cal M}}^\ast(E_{m_2},\omega)
  \Delta E^2 . \nonumber
\end{eqnarray}
For $N\rightarrow \infty$ the second term
vanishes. Taking the limit $\Delta E\rightarrow 0$ we finally obtain
\begin{equation}
  {\cal E}(|\widehat{{\cal M}}_{fl}(\omega)|^2) = A_pt_{obs}
  \int_{E_{min}}^{E_{max}} 
                f(E) |\widehat{{\cal M}}(E,\omega)|^2 dE.
\end{equation}

White noise adds in the following way. Let the total emission measure with
noise be
\begin{equation}
  {\cal M}_{tot}(t) = {\cal M}_{fl}(t) + {\cal M}_{noise}(t)
\end{equation}
where
\begin{equation}
  {\cal M}_{noise}(t) = r d{\cal M}
\end{equation}
with $r$ a random variable between $-0.5$ and $+0.5$ and $d{\cal M}$ the
noise interval. Then
\begin{eqnarray}
  \widehat{\cal M}_{tot}(\omega) &=& 
  \widehat{\cal M}_{fl}(\omega) + r \widehat{d{\cal M}} \\
  |\widehat{\cal M}_{tot}|^2 &=& 
  |\widehat{\cal M}_{fl}|^2 + r^2 |\widehat{d{\cal M}}|^2 + \nonumber\\
  && + r \left( \widehat{\cal M}_{fl}\widehat{d{\cal M}}^\ast
    + \widehat{\cal M}_{fl}^\ast\widehat{d{\cal M}} \right) \\
  {\cal E}\left(|\widehat{\cal M}_{tot}|^2\right) &=&
  {\cal E}\left(|\widehat{\cal M}_{fl}|^2\right) +
  {\cal E}(r^2)|\widehat{d{\cal M}}|^2 + \nonumber\\
  && +{\cal E}(r)\left( \widehat{\cal M}_{fl}\widehat{d{\cal M}}^\ast
    + \widehat{\cal M}_{fl}^\ast\widehat{d{\cal M}} \right).
\end{eqnarray}
It is ${\cal E}(r)=0$, ${\cal E}(r^2)=1/12$ and
\begin{eqnarray}
  |\widehat{d{\cal M}}|^2 &=& \sum_{i,j=1}^{N_{obs}}
  (d{\cal M})^2 {\rm e}^{-i\omega(t_i-t_j)}\Delta t^2 \\
  &=& \sum_{i=1}^{N_{obs}}(d{\cal M})^2 \Delta t^2 \\
  &=& t_{obs} \Delta t(d{\cal M})^2.
\end{eqnarray}
Thus,
\[
  {\cal E}\left(|\widehat{\cal M}_{tot}(\omega)|^2\right)=
\]
\begin{equation}
  =  t_{obs}\cdot \left[ A_p\int 
                f(E) |\widehat{{\cal M}}(E,\omega)|^2 dE
                + \frac{\Delta t}{12}(d{\cal M})^2 \right].
\end{equation}


\bibliographystyle{apj}
\bibliography{1072Kraev}

\end{document}